\documentclass[twocolumn,showpacs,preprintnumbers,amsmath,amssymb]{revtex4}

\usepackage{graphicx}
\usepackage{dcolumn}
\usepackage{bm}

\begin{document}

\title{Implications of the B20 Crystal Structure \\
              for the Magneto-electronic Structure of MnSi}

\author{T. Jeong}
 
\author{W. E. Pickett}
 
\affiliation{
Department of Physics, University of California, Davis, California 95616
}

\date{\today}

\begin{abstract}
Due to increased interest in the unusual magnetic and transport behavior 
of MnSi and its possible relation to its crystal structure (B20) which
has unusual coordination and
lacks inversion symmetry, we provide a detailed analysis of the electronic
and magnetic structure of MnSi. 
The non-symmorphic {\it P2$_1$3} spacegroup leads to unusual fourfold degenerate
states at the zone corner R point, as well as ``sticking'' of pairs of
bands throughout the entire
Brillouin zone surface.  The resulting Fermi surface acquires unusual
features as a result of the band sticking.  For the ferromagnetic 
system (neglecting the
long wavelength spin spiral) with the observed moment of 0.4 $\mu_B$/Mn,
one of the fourfold levels at R in the minority bands falls at the Fermi energy
(E$_F$), and a threefold majority level at $k$=0 also falls at E$_F$.
The band sticking and presence of bands with vanishing velocity at E$_F$
imply an unusually large phase space for long wavelength, low energy
interband transitions that will be important for understanding
the unusual resistivity and
far infrared optical behavior.
\end{abstract}

\pacs{Valid PACS appear here}

\maketitle

\section{\label{sec:level1}Introduction}

Although the binary compound MnSi has been of interest for some
time, unusual behavior of this material has led to accumulating evidence
that it is a novel type of weak ferromagnet.  
MnSi shares a low symmetry (B20) structure with the monosilicides
of Cr, Fe, and Co that is particularly interesting because of its
lack of a center of inversion.  Isostructural FeSi has attracted substantial 
interest as a correlated electron insulator (sometimes referred to
as ``Kondo insulator'') and although the gap of 0.13 eV is 
reproduced very well by usual density functional based (unpolarized)
band theory as shown by Mattheiss and Hamann, (MH)\cite{mattheiss} 
a local moment is 
evident at elevated temperature and
correlation effects are obvious.
The existence of the gap in band theory seems to be a rather 
delicate one depending strongly on the internal structural parameters
and Fe-Si hybridization, yet the gap persists in (paramagnetic) MnSi 
where it lies above the gap. Since transition metal silicides (CrSi and
CoSi, besides the two mentioned) with varying electron concentrations
take this structure, the gap cannot be instrumental in stabilizing
this structure.

While FeSi remains paramagnetic in spite of displaying local moment
behavior, MnSi becomes magnetically ordered
below 29 K.  At higher temperatures its susceptibility is Curie-Weiss
like, with a Mn moment of 2.2 $\mu_B$.\cite{wernick}  It was established by 
Ishikawa {\it et al.}\cite{ishikawa}
that the order is that of
a long-wavelength heliomagnet\cite{shirane} (wavelength $2\pi/q
\approx 190$ \AA, $q \approx \frac{1}{20}\frac{\pi}{a}$) with an ordered 
moment of 0.4 $\mu_B$/Mn.  The spiral structure has been 
attributed to the lack of inversion symmetry in its B20 crystal
structure, which brings the Dzyaloshinski-Moriya interaction into
play.\cite{bak}  The Curie temperature drops with pressure
until magnetic order disappears\cite{thessieu,lonz1}
at the quantum critical point (QCP) of
$P_c =1.46$ GPa, a modest pressure that corresponds to
a rather small volume change from ambient pressure.  
A magnetic field of 1 kOe leads to
a conical ordered phase, while 6 kOe is sufficient
to transform the system to ferromagnetic (FM) order.\cite{thessieu3} 

The resistivity $\rho(T)$ under pressure and field  has attracted considerable
attention.  Thessieu {\it et al.} reported magnetoresistance that showed
strong structure in the 0.5-1.2 T range for pressures  of 0.7-1.7
GPa.\cite{thessieu} 
Near the critical pressure $P_c$,
they observed $\rho \propto T^{1.7}$ behavior
in zero field that reverted to $T^2$
at 3.3 T. 
Initially
the resistivity $\rho$, which is Fermi-liquid-like ($T^2$) in the
FM phase, was reported to be linear-in-T for 
$P > P_c$,\cite{lonz1} but recent data show it to be $\rho 
\propto T^{3/2}$ up to 30 kbar (twice $P_c$).\cite{lonz2}
This scaling suggests something different from both normal Fermi liquid
behavior ($\rho \propto T^2$) and the T-linear non-Fermi
liquid dependence seen in several metals near the quantum
critical point, and non-universal. Recent zero-field $^{29}$Si NMR 
data\cite{yu} have been interpreted in terms of an 
onset of inhomogeneous magnetism
at 1.2 GPa which extends at least to 1.75 GPa, completely encompassing
the QCP.  Earlier Thessieu {\it et al.} had reported,\cite{thessieu2}
also from $^{29}$Si NMR
studies, that some kind of local magnetic order remained above 
P$_c$.  
For an inhomogeneous phase, a non-Fermi-liquid $\rho(T)$
may not be so meaningful.

There are other data that suggest unusual correlated electron
behavior (magnetic fluctuations, presumably)
for MnSi.  The field dependence of the muon spin 
relaxation rate reported by Gat-Malureanu {\it et al.}\cite{gat}
requires an unconventional explanation.  In addition, the
optical conductivity even at ambient pressure
is strongly non-Drude-like, with the 
behavior having been represented by Mena {\it et al.} in terms
of a frequency-dependent effective mass.\cite{mena}  The 
complex low frequency optical conductivity can be modeled by the
form $\sqrt{\gamma(T) + i\omega}$, where $\gamma(T)$ is the
T-dependent relaxation rate.
The antisymmetric part of the magnetic susceptibility that appears
due to the non-centrosymmetric space group has recently been measured
by Roessli {\it et al.}\cite{roessli}  The fluctuations well above
T$_c$ are incommensurate with the lattice, and furthermore are chiral
in nature, meaning that the fluctuation 
spectrum depends on the sign of the 
polarization with respect to the momentum transfer ($\vec P_{pol}\cdot
\vec Q$).

\begin{figure}
\vskip -5mm
\includegraphics[width=8cm,angle=0]{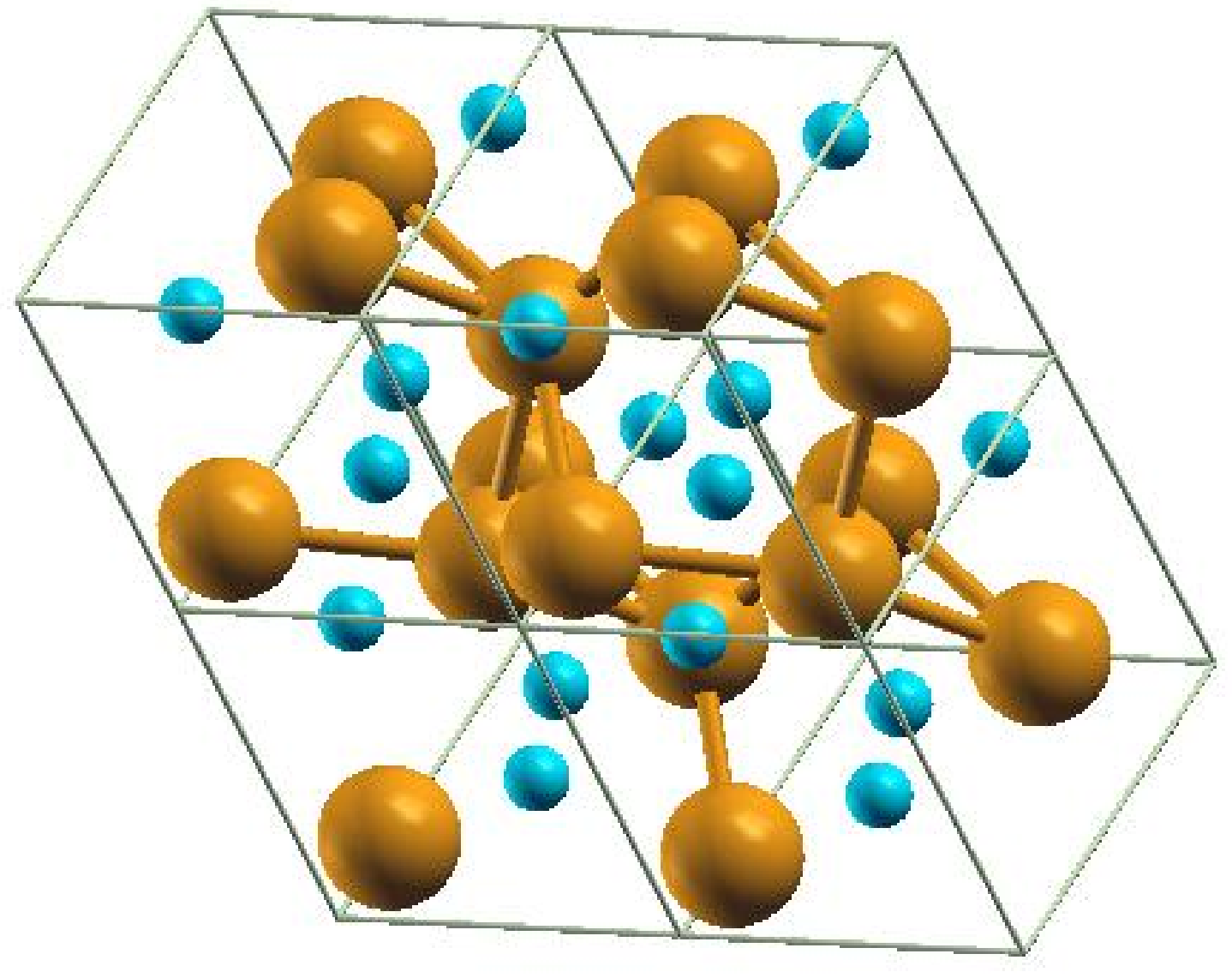}
\vskip -15mm
\includegraphics[width=8cm,angle=0]{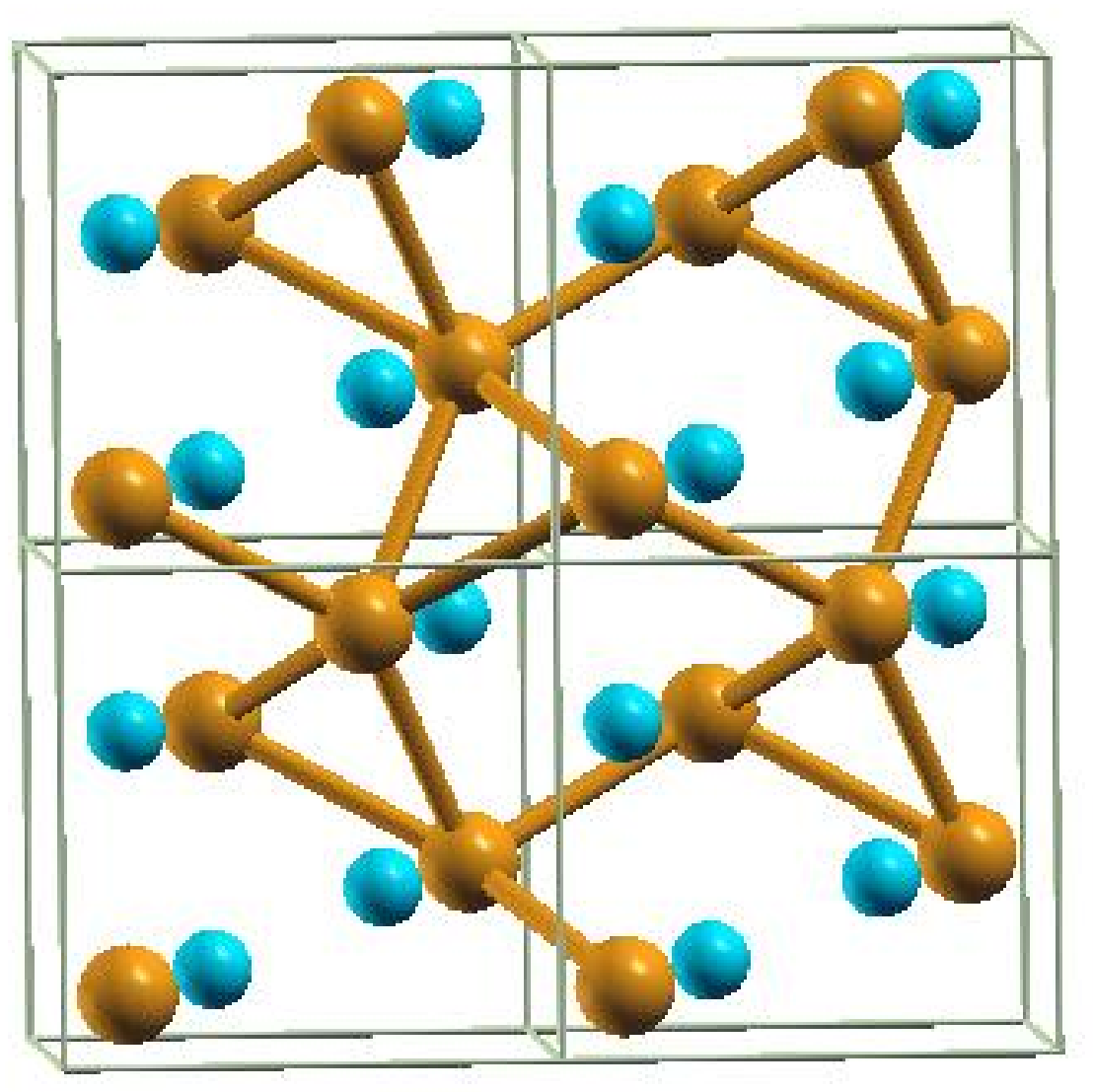}
\caption{(Color online)
Two views of the B20 crystal structure of MnSi, showing four cells. 
The larger atoms are Mn and are connected by sticks; the smaller spheres
are Si atoms. 
Top: a view along the (111) direction.  Bottom: view nearly along the 
(100) axis.} 
\label{structure}
\end{figure}

As mentioned above,
MnSi shares the B20 crystal structure with FeSi, but contains
one electron less per formula unit.   
In spite of the strong interest in this unique compound, there
has been no thorough study of its electronic and magnetic 
structure in general, and in particular how it relates to 
the B20 crystal structure.  The paramagnetic bands and density of states
were presented in a short report by Nakanishi {\it et al.}\cite{nakanishi}
Taillefer and collaborators applied the
linear muffin-tin orbital method with band shifting to model the
Fermi surfaces,\cite{strange} but did not report the 
general electronic structure and
magnetic behavior.  Trend studies of several transition metal monosilicides
using a planewave pseudopotential method were reported by Imai 
{\it et al.}\cite{imai}  Yamada and coworkers have reported studies of
the volume dependence of the magnetic moment using the linear muffin-tin 
orbital method in the atomic sphere approximation.\cite{yamada1,yamada2}

In this paper we present a detailed investigation of the electronic 
structure and magnetization of MnSi within the local density approximation
(LDA) using self-consistent full potential methods.  We address specifically
the connection of the electronic and magnetic structure to the unusual
coordination of the Mn atom and to the non-symmorphic B20 space group 
which has no inversion operation.  We find that ``band sticking'' around
the faces of the Brillouin zone have consequences for the electronic
structure and Fermi surfaces that should be recognized when interpreting
data.  The LDA minimum of energy occurs for a ferromagnetic moment of
almost 1 $\mu_B$/Mn, considerably higher than the experimental value of
0.4 $\mu_B$ and also quite different from the value of $0.25 \mu_B$ 
reported earlier by Taillefer, Lonzarich, and Strange\cite{strange} with
a more approximate method of calculation.  Fixing the moment at the 
observed value, we report the resulting Fermi surfaces.  In the minority
band, considerable potential for nesting arises.   The manner in which
these features may influence the observed properties of MnSi are discussed. 

\section{The B20 Crystal Structure}
One view of the MnSi B20 structure is shown in Fig. \ref{structure} 
There are four 
MnSi formula units in the primitive cell.
MH have described in detail how the local coordination 
of an individual Mn or Si atom can be pictured
in terms of an underlying rocksalt unit cube containing four formula units.
Starting from the rocksalt idealization,
one considers a dimerizing-type distortion involving displacement of 
Mn or Si atoms along  $[111]$ 
directions, whose primary effect is to transform 
rocksalt Mn-Si neighbors along $[111]$ directions into MnSi nearest neighbors. 
These distortions are large (1.6 eV energy gain
per molecule calculated by MH), and 
strongly distort the space-group symmetry from {\it Fm3m} to 
{\it P2$_1$3}, a primitive lattice generated by a screw axis {\it 2$_1$} and
a threefold axis ({\it 3}).  The Bravais 
lattice remains simple cubic but the point symmetry is reduced to four
threefold axes.  The 
space group consists of threefold rotations around 
one specific $<111>$ direction, three screw axes consisting of a 180$^{\circ}$
rotation around a cubic axis followed by a $(\frac{1}{2},\frac{1}{2},0)a$
type non-primitive translation, and combinations of these.  

Both the Mn and Si atoms are located at the 4(a)-type sites in the simple-cubic unit cell, with position coordinates at  
$(u,u,u), (\frac{1}{2}+u,\frac{1}{2}-u,-u), (-u, \frac{1}{2}+u,
\frac{1}{2}-u),$
and $(\frac{1}{2}-u, -u, \frac{1}{2}+u)$. 
The corresponding values for the internal atom-position parameters are
$u_{Mn} = 0.137 $ and $u_{Si} = 0.845 $. 
Given the MnSi lattice parameter is a=$4.558\AA$,
\cite{nakanishi} the local coordination of Mn consists of one Si neighbor
at 2.11~\AA~(lying along a [111] direction, three neighbors at 2.35~\AA,
and three neighbors at 2.69~\AA.  
The point symmetry at the Mn and Si sites is $C_{3}$.
Taking into account time-reversal symmetry (when non-magnetic) with the
twelve space group operations, 
the irreducible
Brillouin zone (BZ) is 1/24 of the full zone.

Another way to picture the B20 structure has been discussed by 
Vo\v{c}adlo {\it et al.}\cite{vocadlo} in terms of an ``ideal B20''
structure with $u_{Mn}=1/(4\tau) = 0.1545$, $u_{Si}=1 - 1/(4\tau) = 0.8455$,  
where $\tau = (1+\sqrt{5})/2$ is the golden ratio.  The nearest neighbor
coordination of each atom then is seven equidistant atoms of the opposite
kind, at a distance $a\sqrt{3}/(2\tau)$.  These seven atomic sites lie
on seven of the twenty vertices of a pentagonal dodecahedron centered on
the atom.  This sevenfold coordination also supports the interpretation
of Dmitrienko that the B20 structure can be viewed as a crystalline
approximation to an icosahedral quasicrystal.\cite{dmitrienko}  
To our knowledge, there 
has been no clear interpretation in terms of chemical bonding
of why MnSi (and CrSi, FeSi, CoSi)
prefer the B20 structure.  Vo\v{c}adlo {\it et al.} have calculated the
equation of state for FeSi in the B20, NaCl, CsCl, NiAs, and inverse-NiAs
structures using the VASP planewave code.\cite{VASP}  
They find that under pressure the internal coordinates 
move closer to their ideal values discussed above, and that a transformation
to the CsCl structure is predicted at 13 GPa. (This prediction is at best
an underestimate, since the B20 structure is known to be stable to beyond
P$_c \approx$ 15 GPa.)

\section{Method of Calculations}

We have used the full-potential 
nonorthogonal local-orbital (FPLO) method\cite{FPLO} within the local 
density approximation (LDA).\cite{perdew}  
Mn $3s,3p,4s,4p,3d$ states and Si $3s,3p,3d$ were included as 
valence states. All lower states were treated as core states.
The inclusion of the relatively extended $3s,3p$ semicore states as band states was done because of the considerable overlap of these 
states on nearest neighbors.
This overlap would otherwise be neglected in the FPLO scheme. Si $3d$ 
states were added to increase the quality of the basis set. 
The spatial extension of the 
basis orbitals, controlled by a confining potential $(r/r_{0})^4$, was 
optimized to minimize the total energy. The self-consistent potentials were 
carried out on a $20\times 20\times 20$ uniform mesh in the Brillouin zone,
which 
corresponds to 700 k points in the irreducible part for {\it P2$_1$3}.

\section{Results and Discussions}

\begin{figure}
\includegraphics[height=8.5cm,width=8.5cm,angle=-90]{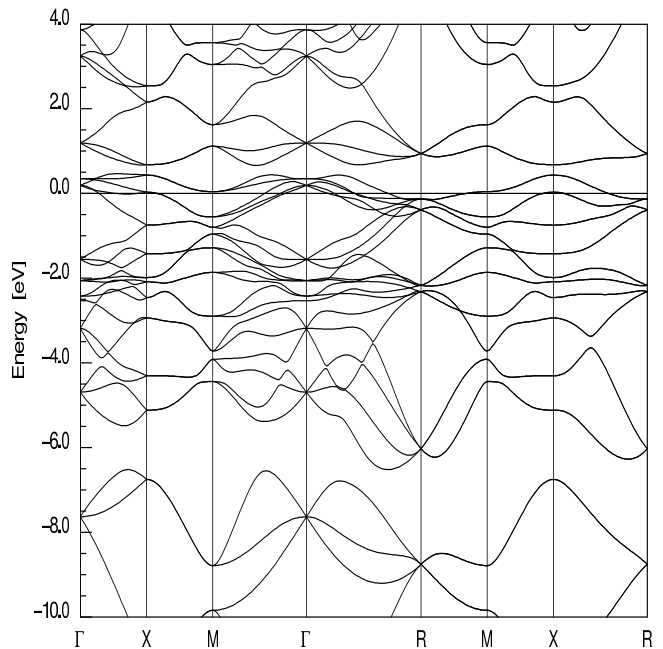}
\caption{The full band structures of paramagnetic MnSi along symmetry lines.
There is a narrow gap in the bands $\sim$0.6 eV
above the Fermi level, corresponding to a band filling of four extra
electrons.  Bands from -3 eV to +1 eV are primarily
Mn $3d$ character.
}
\label{fullbands}
\end{figure}

\begin{figure}
\includegraphics[height=8.5cm,width=8.5cm,angle=-90]{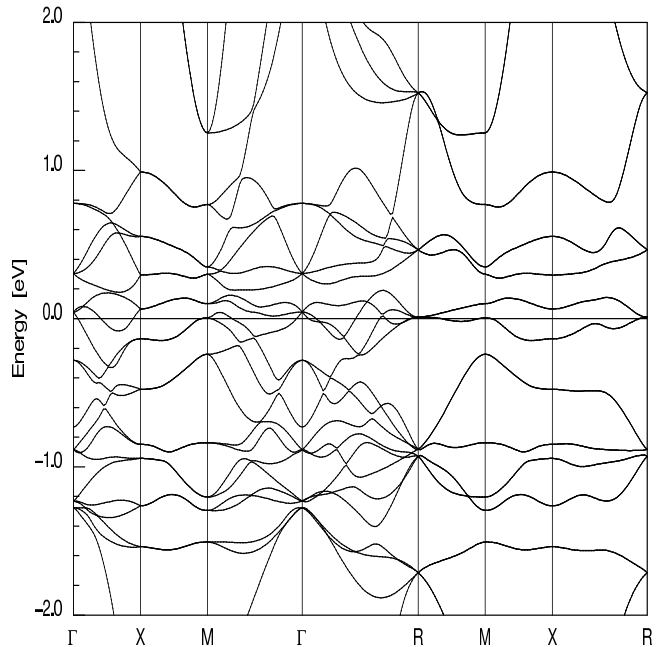}
\caption{Plot along symmetry directions of the bands of Mn$\Box$, that is,
Mn atoms in the observed MnSi B20 structure but with Si atoms missing.  
The amount of
dispersion gives an impression of the ``intrinsic'' $3d$ bandwidth, while
comparison with the full band structure in Fig. \ref{fullbands} indicates the 
considerable shifting of bands due to Si interaction with Mn; for example,
at $\Gamma$ in MnSi, there are no states in the -1.5 eV to 0.3 eV region.
}
\label{boxbands}
\end{figure}

\begin{figure}
\includegraphics[height=8.2cm,width=8.2cm,angle=-90]{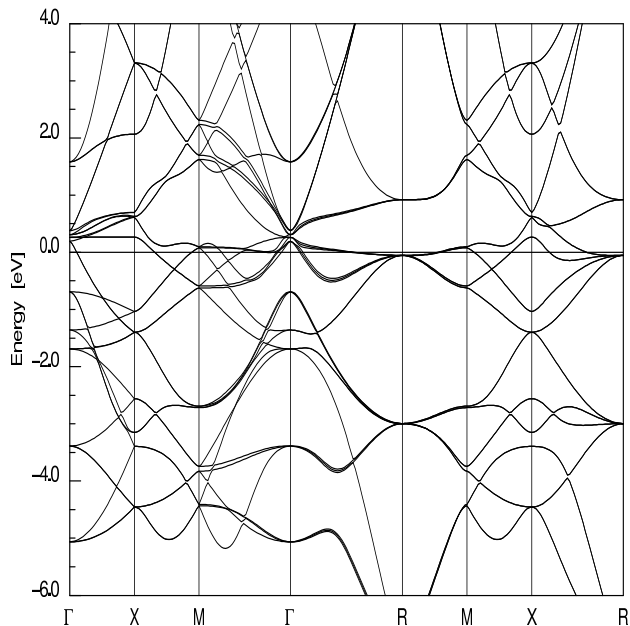}
\caption{Band structure for  
nonmagnetic MnSi in the simple cubic four-molecule non-primitive cell of the
reference rocksalt structure MnSi.  Actually, both Mn and Si have been 
displaced by $\delta u = 0.0025$ to the B20 structure to illustrate lifting
of degeneracies (the tiny band splittings, especially at $\Gamma$ and M)
by the crystal symmetry breaking.  Note that there is no splitting at the
zone corner R point, while fourfold degeneracies at X and M are broken into
``sticking'' pairs. }
\label{rocksalt}
\end{figure}

\subsection{Paramagnetic bands}
Considering the B20 structure as derived from the rocksalt structure,
the MnSi bands look very much like those presented by
MH for FeSi, in both the primitive cell and for the unit cube in which
the X points are folded back to the $\Gamma$ point.  As noted by MH,
there is no reason to expect any Fermi surface driven instability of
the rocksalt structure, because several transition metal monosilicides
with different band fillings have the same structure.  We calculate the
B20 structure to be 1.53 eV per formula unit lower in energy than the
rocksalt structure.

Because of some symmetry-related peculiarities of the band dispersion,
we compare the paramagnetic band structure of MnSi with those of
rocksalt MnSi as well as to those of the Mn skeleton itself.

{\it Full bands of MnSi.}
We first show the full bands of MnSi in the valence-conduction region 
in Fig. \ref{fullbands}.  It is this PM band structure (as slightly
reduced volume) that becomes relevant at the quantum critical
point.  Since Si contributes four, and Mn contributes
seven, valence electrons, there are enough electrons to fill 
11$\times$4 molecules/2 spins
= 22 bands.  With the four $3s3p$ states of Si and the five $3d$ states of
Mn, there are 36 bands valence-conduction bands.  The lowest four 
bands in the range -6.5 eV to -11 eV in Fig. \ref{fullbands} 
are Si $3s$ bands.
Mn $3d$ character is 
confined primarily to the -3 eV to +1 eV region, 
nevertheless these bands are strongly affected 
by the mixing with Si (see below).  As a result of the Mn $d$ - Si $p$
interaction, the Si $3p$ character is repelled strongly to the
region -3 eV to -6 eV in the valence bands, and above 1 eV in the conduction
bands.

A very narrow indirect gap of 0.1 eV lies just above the Fermi level.
This gap, which is the fundamental gap in semiconducting FeSi,
occurs above the MnSi Fermi level since this system contains four
fewer
valence electrons per unit cell.  This gap acquires some significance in
establishing the stability of the FM state, at least within LDA (see below). 
  
{\it Mn$\Box$.}  To give an impression of Mn bands undisturbed by Si
$p$ states, and therefore an indication of direct Mn $d-d$ coupling, we
show in Fig. \ref{boxbands} the bands of Mn$\Box \equiv$ 
MnSi with the Si atoms missing.
The $3d$ bands are about 2 eV wide at $k$=0 (the density of states extends
over 2.5 eV).  As is the case also for MnSi itself (Fig. \ref{fullbands}),
bands ``stick together'' in pairs 
along the BZ edges R-X-M-R.  This symmetry-dictated band-sticking and
hybridization make the dispersion appear small along the BZ faces, but
the bands emanating radially (along $\Gamma$-X and $\Gamma$-R directions)
reflect the actual dispersion.
At $\Gamma$ the twenty states (four atoms $\times$ five $3d$ states) 
break up into a single nondegenerate
state, two twofold degenerate, and five threefold degenerate states.
At the zone corner R point, all bands show unusual fourfold degeneracy,
again a consequence of the non-symmorphic nature of the B20 space group. 

The zone-edge bands show one other unusual feature.  The twenty bands
break up into a single sticking pair at the bottom and the top, and
into {\it pairs} of sticking pairs from -1.5 eV to
0.5 eV that are disjoint all around the Brillouin zone boundary.   
Interaction with the Si $3p$ states completely destroys this simplicity,
(Fig. \ref{fullbands}) even though rather little Si $3p$ character 
remains in these bands.

{\it Rocksalt MnSi.}  For MnSi in the reference rocksalt structure, 
the bands in the -3 eV to 2 eV range are shown in
Fig. \ref{rocksalt}. The Mn and Si positions have been displaced
slightly from the rocksalt
structure toward the B20 structure: instead of using $u_{Mn}=0.25, u_{Si}
=0.75$ which would be rocksalt, the values $u_{Mn}$=0.2475, $u_{Si}$=
0.7475 were used to give some impression of the band splittings
that occur for small distortion toward the B20 structure. (This 
distortion is only about 2\% of that needed to produce the actual B20
structure.
The MnSi B20 ``distortion'' is so large that the
broken symmetries cannot be located just from the B20 band structure.)

A very flat band around the R point lies almost at E$_F$, and flat bands lie 
0.3 eV above E$_F$ along $\Gamma$-X and very near E$_F$ along $\Gamma$-M.  
The occurrence of flat bands is not uncommon is cubic structures, where
certain $d-d$ hopping amplitudes can vanish by symmetry. 
The distortion can be seen to split degeneracies and thereby give vanishing
velocities at the M point, and this does not occur at the X point.  In
the MnSi bands of Fig. \ref{fullbands}, it is evident that bands along
$\Gamma$-X hit the X point with non-zero velocity (in fact, pairs of bands
hit X with equal magnitudes but opposite sign of their velocity), 
whereas all bands
are flat at the M point.

\begin{figure}
\includegraphics[height=8.2cm,width=8.2cm,angle=-90]{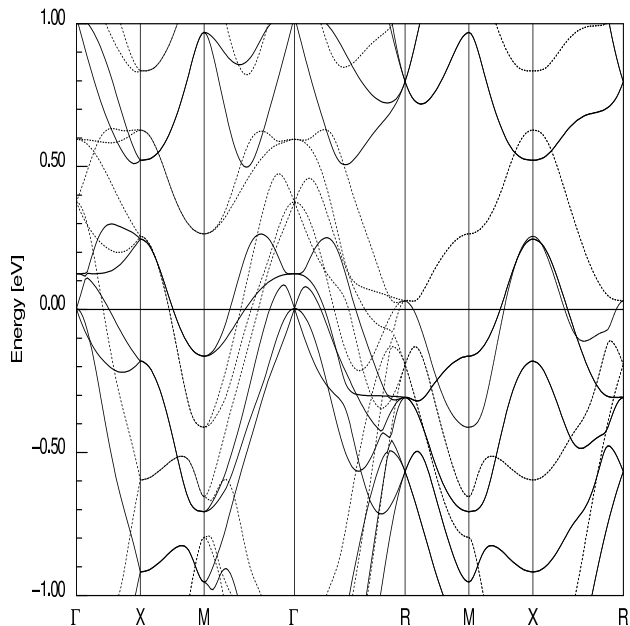}
\caption{The bands of ferromagnetic MnSi within 1 eV of the Fermi level
when the moment is fixed at
the experimental value of 0.4 $\mu_B$/Mn.  The exchange splitting is
fairly uniform at 0.4 eV.  Note that a fourfold degenerate level in
the majority bands falls precisely at the Fermi level, and a fourfold 
minority state at R lies very close to E$_F$.} 
\label{ferrobands}
\end{figure} 

\subsection{Digression on Band Sticking}
To understand better the extensive band sticking noted in the previous
subsection, and because 
we will encounter fine structure in the band structure at E$_F$
in the (physical) FM case in the next subsection, it is worthwhile to
digress briefly to consider the origin of the band sticking phenomenon.
Perhaps the most accessible description of the origins of band sticking can
be found in the recent papers of K\"onig and Mermin.\cite{mermin1,mermin2}
Band degeneracy at a wavevector $\vec k$ is associated with the 
little group of $\vec k$, which is the subgroup of the point group 
operations $\{g\}$ that bring $\vec k$ back to itself,  
each modulo a reciprocal lattice
vector $\vec G_g$.  General (non-symmetric) points on the face of the BZ
of the {\it P2$_1$3} space group have a little group containing only the
identity.  
To obtain the full symmetry of the actual band structure, we have to augment the
symmetry considerations to include time reversal ${\cal T}$, whose effect is
${\cal T} \psi_k(r) = \psi_{-k}(r)$, {\it i.e.} its effect 
is to invert the wavevector ${\cal T}\vec k = -\vec k$ 
just as would an inversion
operation (which is missing from {\it P2$_1$3}).  Thus in the operations
on $\vec k$, we include ${\cal T}$ together with the actual point 
group operations.

Consider a point on the BZ face $\vec K = (\pi,k_y,k_z)$ (taking
unit lattice constant).  Now the little group of $\vec K$ consists of
the identity ${\cal I} = g_1$ and the product ${\cal T}S = g_2$, 
where $S$ is the 180$^{\circ}$ rotation around the $\hat x$ axis
$(x,y,z) \rightarrow (x,-y,-z)$.  When combined with the
non-primitive translation $\tau =(\frac{1}{2},\frac{1}{2},0)$, it forms
the screw axis space group operation ${\cal S}=[S;\tau_S]$.  
With two members in the little
group together with the phase factor introduced by $\tau$ for the $g_2$
operation, the analysis of K\"onig and Mermin indicates band sticking 
over the entire $k_x=\pi$ face (and thus all faces) of the BZ.  
Explicit calculation
confirms the band sticking.  The extensive fourfold degeneracy at R 
that was pointed out in the previous subsection arises already in the
rocksalt supercell of Fig. \ref{rocksalt}.  Extra degeneracies are 
expected when a supercell (non-primitive cell) is chosen; it is a
consequence of the {\it P2$_1$3} symmetry that these degeneracies
remain after the rocksalt$\rightarrow$B20 internal distortion.

At the zone corner R point, the little group consists of all space group
operations, often refereed to as ``full symmetry of the $\Gamma$ point.''
In fact, most of the little group members of the R point
are associated with non-zero
reciprocal lattice vectors $\vec G_g$ (unlike the $\Gamma$ point case), 
and associated sticking leads to more
degeneracy than occurs at the $\Gamma$ point. 

\subsection{Ferromagnetic phase}
Since the long wavelength helical magnetic structure can be considered 
to be locally FM, and 
a field of only 6kOe is sufficient to drive the heliomagnetic structure
to FM order, it is relevant to consider a simple FM ordering.  Since
spin-orbit coupling is small in $3d$ magnets, we neglect it, so the
direction of magnetic polarization is not coupled to the lattice.
Within LDA, a FM state with moment almost at the half metallic value of 
1 $\mu_B$/Mn is obtained: the Fermi level in the minority bands lies
just at the bottom of the gap in the majority bands.  The four holes
compared to FeSi lie entirely in the minority states.  This result
is substantially different from experimental moment of 0.4 $\mu_B$/Mn,
and overestimate that is not uncommon for weak ferromagnets near the quantum
critical point.  It does however mean that analyzing the LDA minimum is 
unfruitful.

To reveal the electronic and magnetic structure of MnSi as fully as
possible, we have constrained the moment to the experimental value of
0.4 $\mu_B$/Mn.  The bands in the $3d$ region are shown in Fig.
\ref{ferrobands}.  The exchange splitting $\Delta_{ex}$ = 0.4 eV is 
rather uniform over the zone, as expected for bands which are of
fairly uniform character (Mn $3d$).  Identifying $\Delta_{ex} = I m$
gives a Stoner constant $I = 1$ eV/$\mu_B$, a value near the top of
the range occurring in $3d$ magnets.

The FM bands are nearly uniformly split versions of the paramagnetic
bands.  The notable feature is the coincidence of two complexes of
degenerate bands at high symmetry points that lie essentially at the
Fermi level.  (Since the magnetization has not been determined to
any more accuracy than ``0.4 $\mu_B$/Mn'' there is an uncertainly of
several tens of meV of band placements.)  At $k$=0, a threefold level
lies precisely at $E_F$.  The middle of these three bands approaches
$k$=0 with vanishing slope, while the other two have non-vanishing
(equal and opposite in sign) velocities of 1.25$\times 10^7$ cm/s.
At the zone corner R point, a fourfold level lies extremely near E$_F$
(about 15 meV above).   These bands all have vanishing velocity at R.

The occurrence of a vanishing velocity at E$_F$ has a variety of possible
consequences which have been discussed at much length.  The most well
known case is the half-filled square lattice with only nearest neighbor
hopping, where a van Hove singularity occurs at the zone edge X = 
($\frac{\pi}{a},\frac{\pi}{a})$ point in two dimensions.  In addition
to causing the van Hove non-analyticity (and peak) in the DOS, this
situation represents the point of change in topology of the Fermi
surface, which includes a ``Lifshitz anomaly of order 2$\frac{1}{2}$''
in thermodynamic properties at low temperature.

\begin{figure}
\includegraphics[height=5cm,angle=-0]{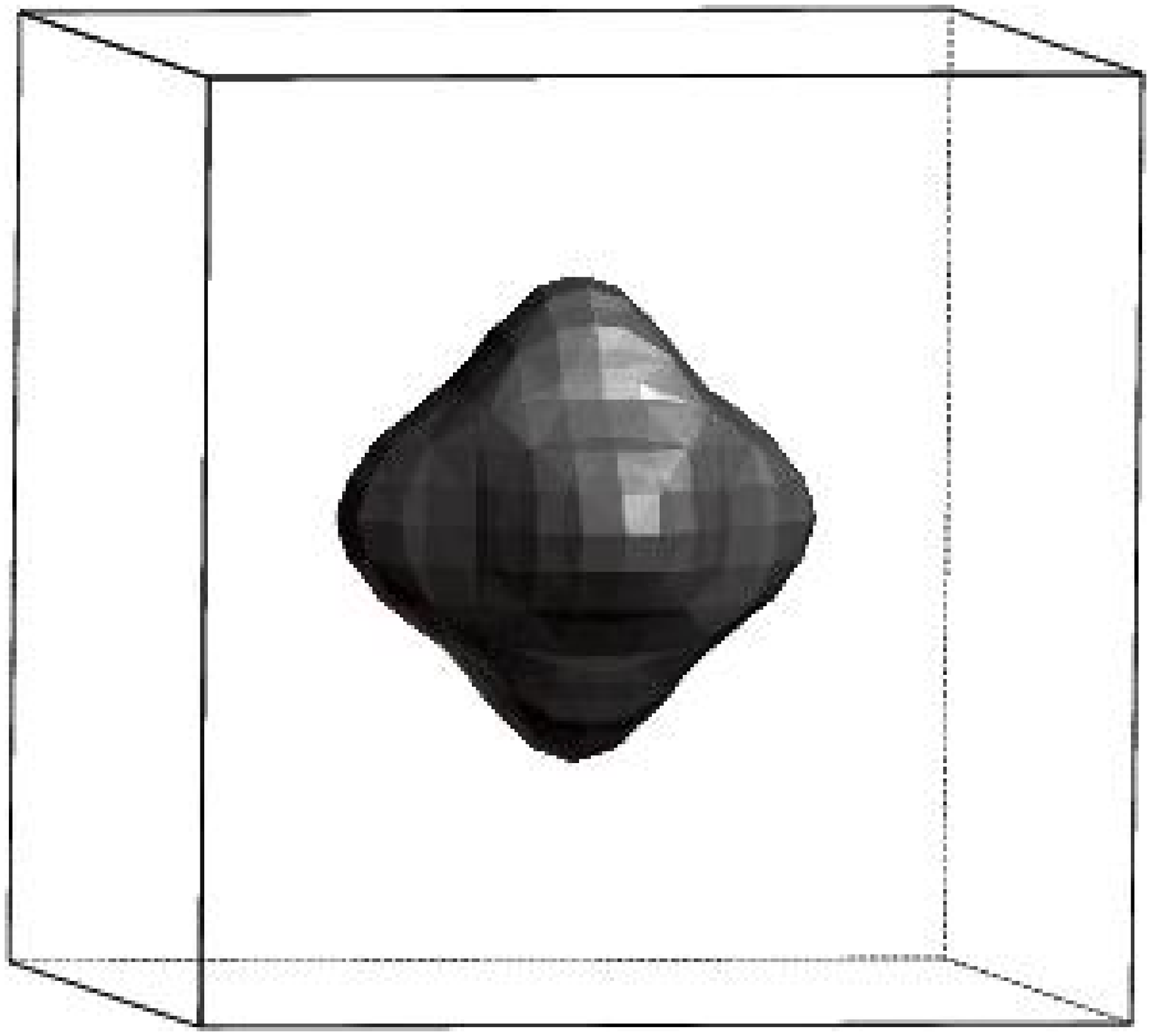}
\includegraphics[height=5cm,angle=-0]{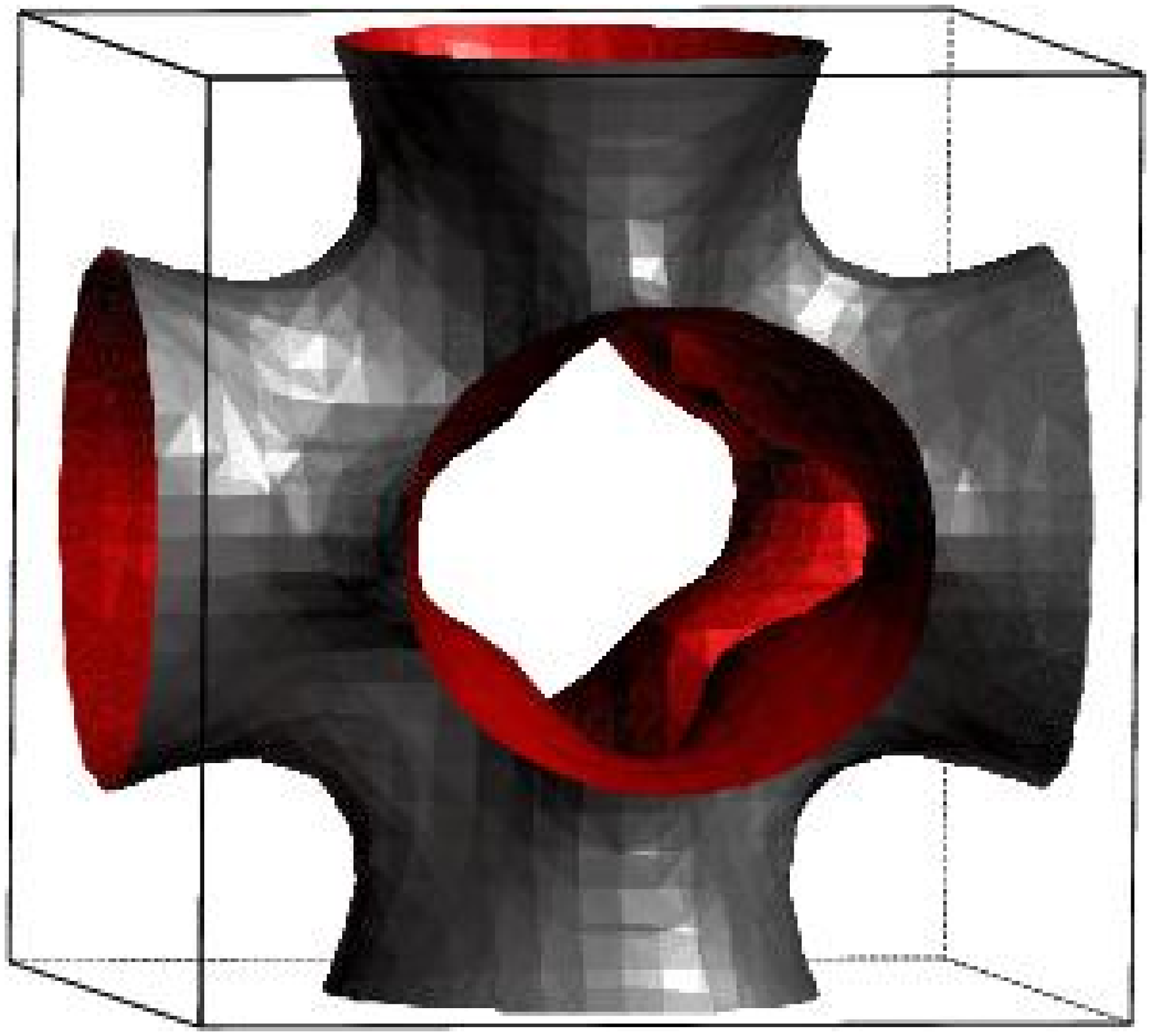}
\includegraphics[height=5cm,angle=-0]{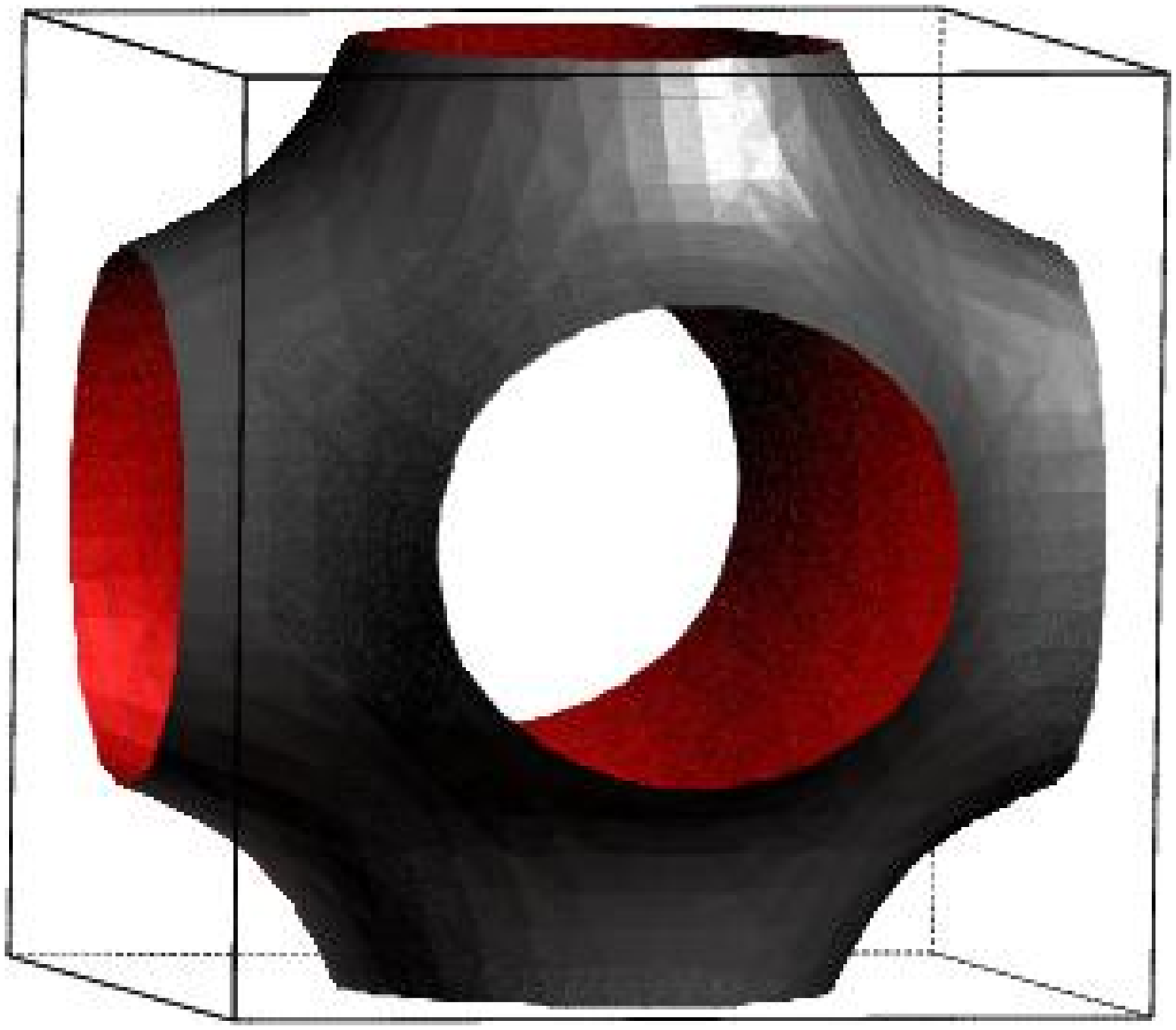}
\caption{(Color online)
Majority spin Fermi surfaces (from top, bands 37, 38, 39) for a fixed
moment equal to the experimental value of 0.4 $\mu_B$/Mn.  The close 
juxtaposition of the lower two panels illustrate how the smooth connection from
one jungle gym surface to the other in a neighboring zone proceeds without
the surfaces being perpendicular to the zone face.}
\label{upFS}
\end{figure}

\begin{figure}
\includegraphics[height=5cm,angle=-0]{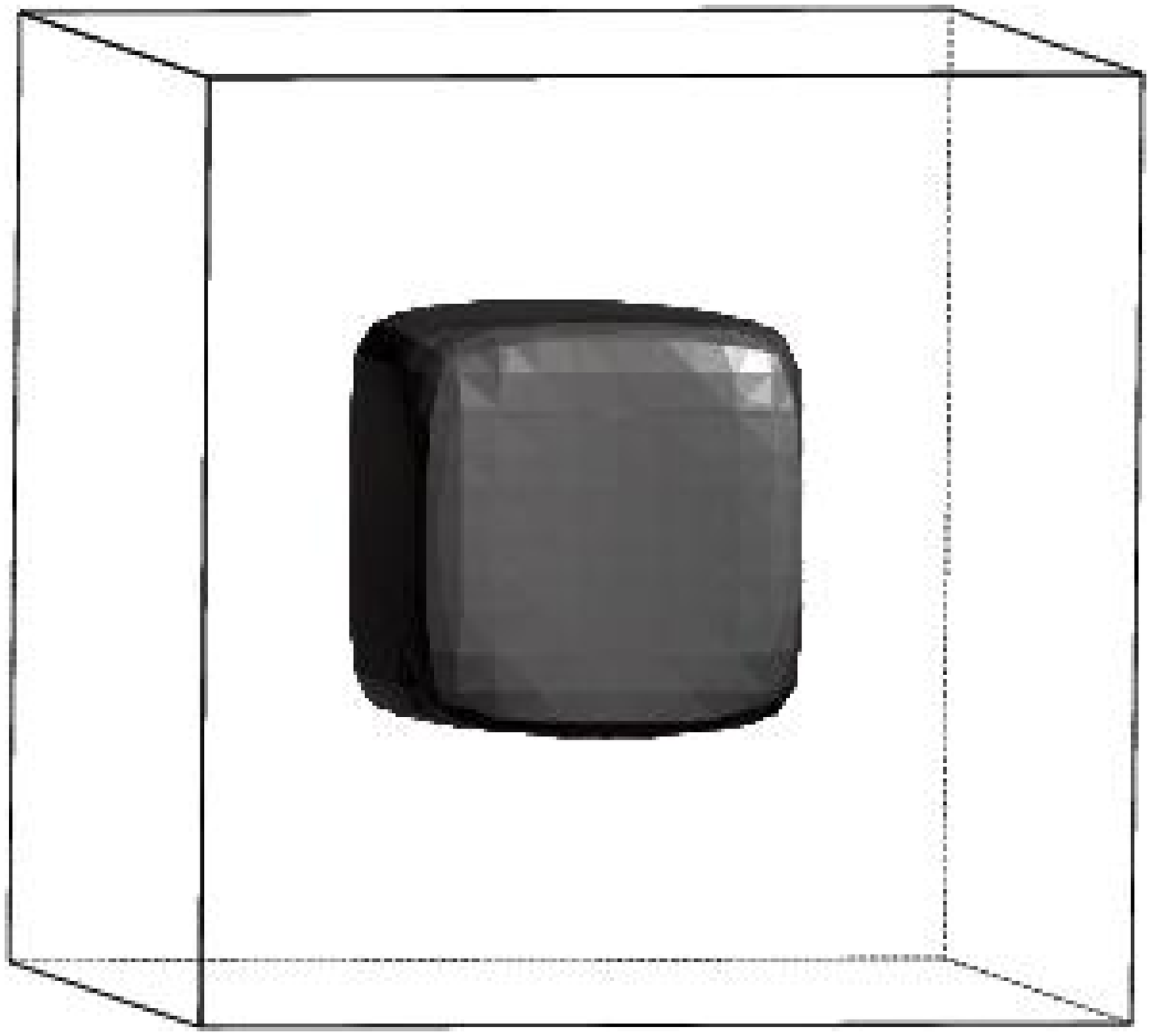}
\includegraphics[height=5cm,angle=-0]{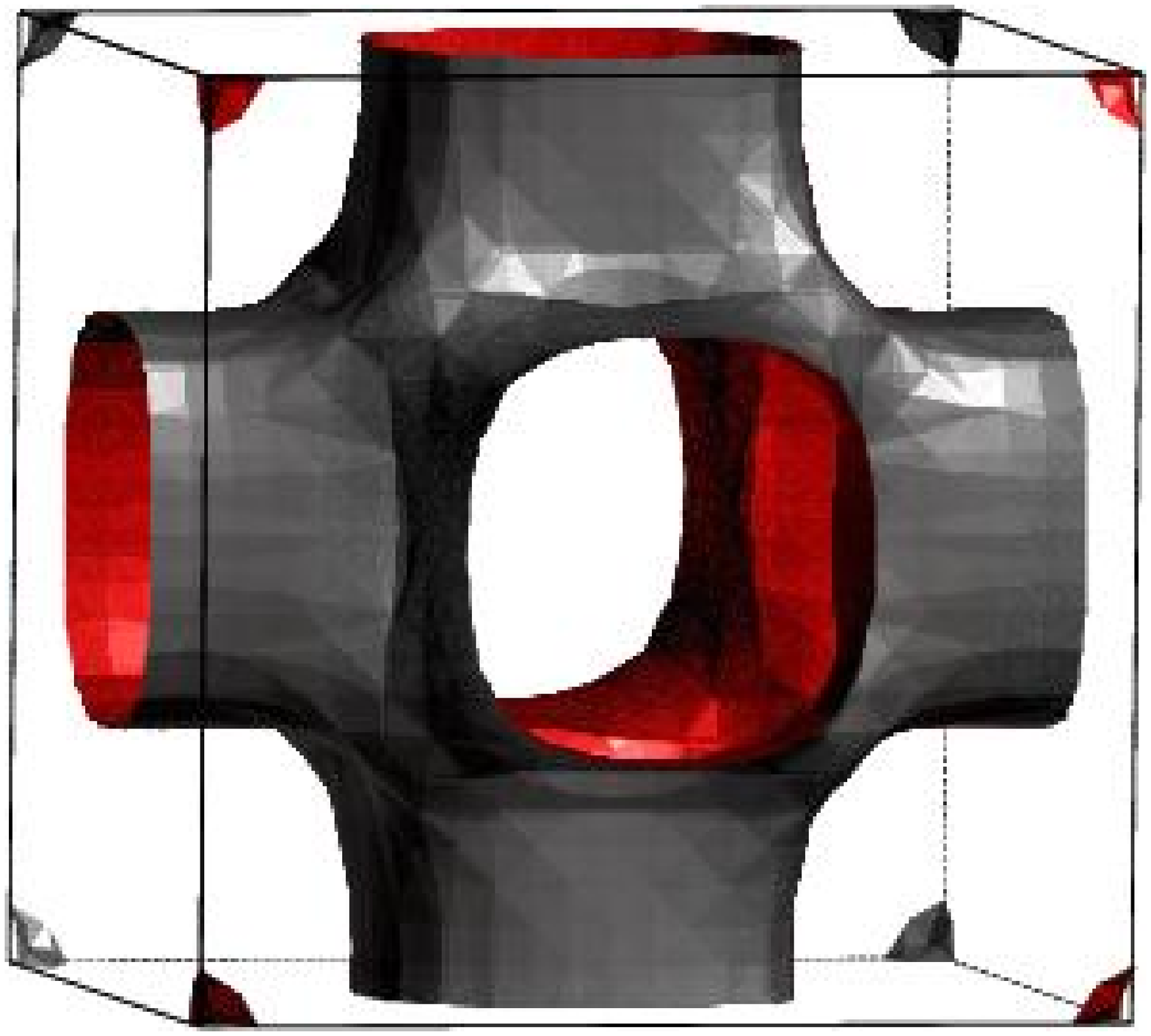}
\includegraphics[height=5cm,angle=-0]{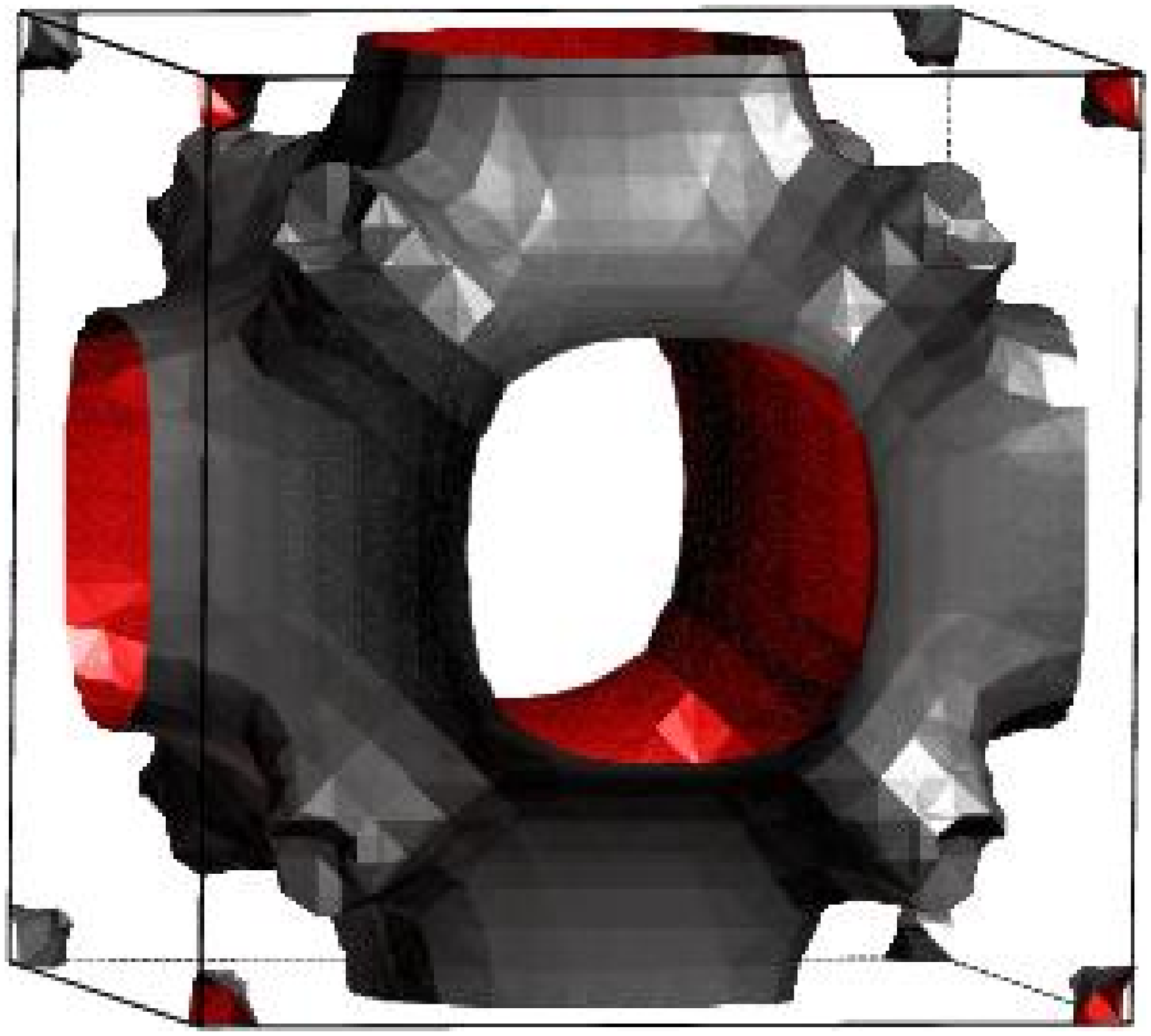}
\caption{(Color online)
Minority spin Fermi surfaces as in Fig. \ref{upFS} (from top, 
band 35, 36, 37).  The strong nesting features (top two panels)
are discussed in the text.
Note that the band 37 jungle gym has additional structured protrusions
along the $<111>$
directions.  Two very small hole ellipsoids lie at the zone corner R point.}
\label{dnFS}
\end{figure}

\subsection{Fermi Surfaces}
The Fermi surfaces for a magnetization of 0.4 $\mu_B$/Mn are shown in 
Fig. \ref{upFS} and \ref{dnFS} for majority and minority spins, respectively.
The majority surfaces consist of a $\Gamma$-centered rounded octahedron and a
pair of open jungle gym type surfaces.  These latter two surfaces illustrate
two consequences of the unusual B20 space group of MnSi.  First, although
the Bravais lattice is simple cubic, the lack of a fourfold rotation 
around the cubic axes is reflected in the elliptical (rather than 
circular) intersection of the arms with the zone faces.  The orientation of
the major axis of this ellipse rotates from face to face in a way that
preserves the threefold rotation symmetry around $<111>$ axes.

More difficult to accommodate with our conception of periodic cells (whether
in real or reciprocal space) is the recognition that these jungle gym arms
do not intersect the zone faces perpendicularly.  The periodicity of the
total Fermi surface is restored by the fact that there are two 
touching Fermi surfaces on these zone faces (in this case, the two jungle
gyms) guaranteed by the sticking together of bands discussed in the 
previous section.  At a given point on the Fermi surface at a zone face,
one surface intersects the face at an angle $\pi/2 -\theta$ and the other
at an angle $\pi/2 +\theta$: each Fermi surface connects smoothly to the
{\it other} Fermi surface in the neighboring zone.  As a result, the open
magneto-oscillation orbits on these surfaces may have behavior different
from what is expected
from the reciprocal lattice periodicity.  In addition,
extremal neck orbits do not automatically encircle the X point as is
commonly the case (see Fig. \ref{upFS}).

The minority Fermi surfaces, shown in Fig. \ref{dnFS} 
consist of 
$\Gamma$-centered cube and two jungle gyms (again, stuck together along
the zone faces) and appear to be direct analogs of the majority surfaces.
There are also two small R-centered ellipsoids that are not well 
resolved in Fig. \ref{dnFS}.
The resemblance of the jungle gyms to those of the majority
band (Fig. \ref{upFS}) is largely accidental.  As mentioned in the
previous section, the spin splitting if nearly uniform at 0.4 eV, so the 
Fermi level produces a cut in two different regions, which are $\pm$0.2
eV from the paramagnetic Fermi level.  The intersections of the jungle
gym surfaces with the zone faces can be seen to arise from majority and
minority bands which have, apparently accidentally, nearly the same shape
along M-X-R in Fig. \ref{ferrobands}.  The similarity in shape of the
jungle gym surfaces (for either spin direction) may be understood simply
by the fact that they are required by symmetry to ``stick'' along each
of the zone faces, and they do not extend too far away from the faces.

The minority surfaces in particular
present a great deal of potential for nesting,
as represented for example in the quantity
\begin{eqnarray}
\xi(\vec q) = \sum_{\vec k} 
  \delta(\varepsilon_{\vec k+\vec q}-\varepsilon_F)
   \delta(\varepsilon_{\vec k}-\varepsilon_F)
\end{eqnarray}
that measures the number of transitions on the Fermi surface involving
momentum transfer $\vec q$.
Taken together, the cube and the more jungle gym in the center panel of
Fig. \ref{dnFS} arise from a set of parallel flat ribbons oriented along
each of the three Cartesian directions to form intersecting square tubes
(somewhat rounded).  The dimension across the tube, hence the nesting
wavevector, is $Q \approx 0.4 \frac{2\pi}{a}$.  Due to the flat nature
of the surfaces, nesting vectors are of the general form $\vec q =
(q_x,Q,0)$ and $\vec q =(q_x,0,Q)$ for arbitrary $q_x$, and others related
by cubic symmetry.  [We say cubic symmetry rather than the strictly
correct ``B20 symmetry'' because the departure of the shapes of the
minority surface from cubic symmetry seems to be less than for the
majority jungle gyms.]

We leave a more detailed study of nesting and $q$-dependence of scattering
processes for subsequent work.  A few other possibilities can be noted, however.
The majority octahedron may provide some nesting  
across the octahedron ``faces'', that is, for $q$ along the $<111>$ 
directions, but they do not appear to be extremely flat.  
In addition, the sticking together of the jungle gym surfaces
all over the zone faces provides substantially larger phase space for
$q\rightarrow 0$ transitions than for conventional Fermi surfaces and
deserves further study. [The trivial {\it intra}band $\frac{1}{q}$
divergence\cite{xiQ} is a different matter, and is usually killed by intraband
matrix elements.]

\section{Discussion}

\begin{figure}
\includegraphics[height=8cm,angle=-90]{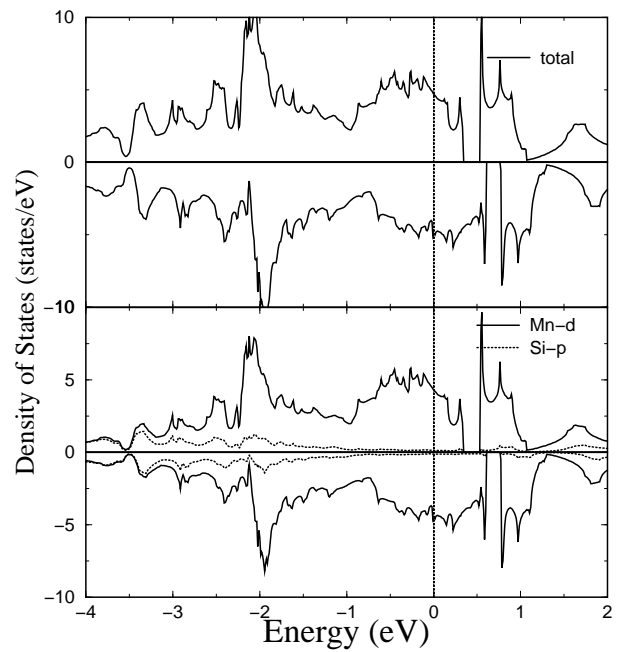}
\caption{Projected density of states of ferromagnetic MnSi.  Top panel:
total, plotted positively for majority, negatively for minority. The near
rigid-band shift of states within 1 eV of the Fermi level is evident.  Bottom
panel: projection of the Mn $3d$ and Si $3p$, showing that 
Mn $3d$ character dominates the states near Fermi level.} 
\label{Proj}
\end{figure} 

The fundamental questions underlying the unusual and perhaps unique 
characteristics of MnSi center on (1) its unusual bonding and coordination,
identical to that of the ``Kondo insulator'' FeSi, which is related to (2)
its low symmetry but still cubic B20 space group in which 75\% of the 
rotations are coupled with a non-primitive translation, and (3) its lack
of inversion center (also an integral part of the space group).
Frigeri {\it et al.} have argued, for example, that
lack of inversion symmetry in the MnSi space group acts as a strong
inhibition\cite{frigeri} to spin triplet pairing (but does not
totally exclude it).
Based on prior study of FeSi, many Mn $3d$ bands 
near the Fermi level should be expected, together with the occurrence of
several valence and conduction band extrema within a few tens of meV of
the Fermi level. 

We have presented a detailed look at the electronic structure of MnSi
both in its paramagnetic phase, which is relevant to the quantum critical
point, and also for the FM phase with the observed value of the
magnetization.  Peculiarities arising from the non-symmorphic nature
of the space group are evident.  One peculiarity is the sticking together 
in pairs of bands over the {\it entire surface} of the Brillouin zone.  This
sticking allows a related uncommon feature: the Fermi surfaces 
contact the BZ face at other than a right angle.  This is done
in pairs, and one result is that there is a much larger phase space
available for $q \rightarrow 0, \omega
\rightarrow 0$ interband transitions than would be the case without
band sticking.  Such transitions
occur only for
intersecting Fermi surfaces; one may say that the B20 space group strongly
encourages intersecting Fermi surfaces (on the BZ faces).  Such 
transitions may be responsible for the unconventional temperature 
dependence of resistivity and frequency dependence of the optical
conductivity.  These questions will be pursued in more detail elsewhere.

The space group also leads to some nonzero velocities at the zone
center as well as on the BZ faces (these latter related
again to the band sticking).
The latter occurrence arises in the threefold degenerate states
at $k$=0: one has vanishing velocity, while the other two have velocities
that are equal in magnitude and different in sign.  This phenomenon 
obtains importance because one such level lies precisely at E$_F$ for
the case of FM order.  Such an occurrence
also marks a point of change of topology of the Fermi surface,
which carries with it in principle a Lifshitz transition of order 
$\frac{5}{2}$.\cite{lifshitz,dagens,wepla,andy} 

There is also the feature of saddle points or other points of vanishing
velocity in the bands falling at E$_F$.
For zero magnetization this occurs at the zone face X point, and also
an M point saddle point is close to E$_F$.  For the
FM case, a vanishing velocity at a fourfold degenerate state at R
lies within 30 meV of E$_F$.  Almost exactly at E$_F$ at the zone center
is a threefold level, where the anomalous feature is not the zero velocity
(usual at $k$=0) but rather the two non-vanishing velocities.  A near
(but not exact) occurrence of non-vanishing velocities at $k=0$ (which
can be interpreted as diverging effective mass) has been studied in 
the skutterudite materials.\cite{skutter}
Such band features definitely influence
the spectrum of low energy excitations of the system, and may help to 
account for the observed peculiarities in the normal state properties
of MnSi.  

The lack of inversion symmetry {\it per se} has not been 
implicated in the peculiarities we have located, partly because 
time-reversal symmetry restores the band symmetry (and band sticking)
that ``normally'' would be provided by inversion.  Affects of the
lack of inversion will arise
in the FM phase when spin-orbit coupling is considered, in which case
(1) some of the band sticking will be relieved, and (2) there is no longer 
enough symmetry to enforce $\varepsilon_{-k} = \varepsilon_{k}.$
One important result is that the Fermi surface becomes ``lopsided,'' 
restricting the formation of zero-momentum superconducting pairs 
built from $\vec k$
and $-\vec k$.  We will pursue this question elsewhere; however,
spin-orbit coupling is small in Mn so its importance will have to be
assessed.  Several systems have shown the appearance of 
superconductivity near the
magnetic quantum critical point, prompting Frigeri {\it et al.} to argue that
lack of inversion symmetry in the MnSi space group acts as a strong
inhibition\cite{frigeri} to spin triplet pairing, but does not
totally exclude it.  Their symmetry considerations are relevant specifically
to the pairing question and are mostly separate (additional) to our
discussion.

\section{Conclusions}
MnSi is attracting great attention due to its quantum critical point under
modest pressure and to the unusual behavior of several normal state properties.
We have provided a detailed analysis of the effects of the B20 crystal
structure -- both the non-symmorphic space group without a center of inversion
and the unusual coordination -- on the band structure of both the paramagnetic
and ferromagnetic phases.  Several types of unusual occurrences have been
identified, and some of them are likely to be implicated in the anomalous
normal state behavior and possibly the lack of any superconducting phase 
in the vicinity of the quantum critical point.

\section{Acknowledgments}
We acknowledge illuminating discussions with J. Kune\v{s}, I. I. Mazin,
and C. Pfleiderer, and important technical assistance from D. Kasinathan
and K.-W. Lee.  W.E.P. acknowledges illuminating
communication on the band sticking 
question with J. Kune\v{s} and N. D. Mermin.
This work was supported by DOE grant DE-FG03-01ER45876.

\end{document}